\documentclass[mlabstract,onecolumn]{jmlr}
\jmlrproceedings{NeurReps}{NeurIPS 2026 Workshop -- Extended Abstract Track}

\usepackage{booktabs}

\jmlrvolume{}
\firstpageno{1}

\jmlryear{2026}
\jmlrworkshop{Symmetry and Geometry in Neural Representations}

\title[Individual Structure Beyond Conversational Role]{Contrastive Neural
Embeddings Reveal Individual Traits Beyond Conversational Role}

\author{\Name{Hubert Huang} \Email{huberth@stanford.edu}\\
  \addr Stanford University, Palo Alto, CA 94305, USA
  \AND
  \Name{Michelle McCleod} \Email{mlmccleod@crimson.ua.edu}\\
  \addr University of Alabama, Tuscaloosa, AL 35401, USA
  \AND
  \Name{Brendan Ames} \Email{B.Ames@soton.ac.uk}\\
  \addr University of Southampton, Southampton SO17 1BJ, UK
  \AND
  \Name{Evie A. Malaia} \Email{eamalaia@ua.edu}\\
  \addr University of Alabama, Tuscaloosa, AL 35401, USA
}

\begin{document}

\maketitle

\begin{abstract}
Contrastive representation learning is increasingly used to recover
low-dimensional structure from neural recordings, but its output is
typically validated by decoding accuracy rather than by the geometry of
the manifold it produces. We apply CEBRA to EEG recorded from
dyads in conversation, and analyze the resulting embedding, which training constrains
to the 2D sphere. Labels describing the dyads, including the absolute
difference between partners' autism-quotient scores, decode well above
chance ($0.77$ against a $0.55$ majority baseline for binary AQ magnitude;
$0.44$ against $0.25$ for the six-class $|\Delta\mathrm{AQ}|$ partition).
However, the two permutation controls have notable differences in results:
permuting labels over a frozen embedding yields $p=0.001$, whereas
retraining the encoder under each permutation yields $p=0.50$. Only the
latter tests the label rather than the geometry. Consistent with this,
spherical mixture structure and per-class dispersion track
identity rather than autism trait differences in dyads; frequency-band and non-oscillatory activity
ablation controls do not change the results. However, participant-level model does separate from
its identity-aware null ($p=0.0099$) while speaker-versus-listener role
analysis performs at chance in the same embedding, indicating a manifold organized by
individual --- and, in contrast with current neurolinguistics models,
almost invariant to speaking vs.\ listening.
Based on these results, we suggest that retraining-based nulls
should be the default for grouped-data contrastive embeddings.
\end{abstract}

\begin{keywords}
contrastive learning, representational geometry, CEBRA, EEG, dyadic
interaction, permutation controls, autism-spectrum quotient
\end{keywords}

\section{Introduction}
\label{sec:intro}
Contrastive methods such as CEBRA \citep{schneider2023learnable} learn from
high-dimensional time series by projecting task-relevant features into a
low-dimensional latent space. Here we analyse CEBRA-embedded dyadic EEG in terms
of the geometry and grouping structure of the resulting manifold, in order to
separate the contributions of neurotype, activity (speaking vs.\ listening), and
neurotype similarity between partners. Because the objective uses cosine
similarity, the embedding is constrained to $S^{2}$, and every question we ask
is a question about angular structure on a sphere.

Recent behavioural work on double empathy argues that communication difficulty
between autistic and non-autistic people reflects a mismatch between
interlocutors rather than a deficit in either one
\citep{crompton2025information}. Neural data from social interaction is a
natural test bed: the unit of analysis is the dyad, and the relevant
labels (i.e. the difference in AQ-10 scores between partners) are properties
of the interacting pair rather than of either brain
\citep{crompton2025information}. Hyperscanning designs carry known inferential
problems \citep{hamilton2021hyperscanning}; we identify one specific to
representation learning.

A label that is constant within a dyad is a proxy for dyad identity. An encoder
that learns which dyad a segment came from will decode that label above chance
without having represented anything about the labelled trait; deep models
readily recover stable individual and group attributes from brain data
\citep{ryali2024deep,ford2021classification}. Distinguishing the two requires
asking whether comparable geometry arises under arbitrary regroupings. We show
that it does not. We then show that a participant-level AQ label, which is not
nested inside the recording group in the same way, does survive an
identity-aware control, and that the geometry it induces is close to invariant
under the speaker/listener exchange.

\section{Data and embedding}
\label{sec:data}

Each dyad contributes two participants and four role-specific EEG recordings
(A~speaking, A~listening, B~speaking, B~listening), exported as time-locked EDF
files cropped to a common onset and offset. 
    Each dyad yields two AQ-10 scores \citep{lundin2019aq10,baroncohen2001aq}. The
dyad-level label is the absolute difference
$|\Delta\mathrm{AQ}|=|\mathrm{AQ}_{\mathrm{sp}}-\mathrm{AQ}_{\mathrm{ln}}|$,
binarised as \emph{AQ magnitude} (low: $\le 1$; high: $\ge 2$). Crucially,
$|\Delta\mathrm{AQ}|$ is constant across every sample and both role mappings of
a dyad---it is a coarsening of the dyad-identity partition. The
participant-level label is that individual's own AQ-10 score, which varies
within a dyad.
    We train CEBRA with a time-delta conditional objective on channel-wise
normalised inputs, using the \texttt{offset10} receptive-field model and cosine
similarity to produce three-dimensional embeddings on $S^{2}$. Quality is
summarised by final InfoNCE loss, goodness of fit, silhouette score, and
$k$-nearest-neighbour decoding ($k=5$). Every accuracy reported below uses
grouped leave-one-dyad-out cross-validation: both role mappings of a held-out
dyad are withheld together.

\section{Decoding results}
\label{sec:decoding}

Dyad-level AQ magnitude decodes well above the majority baseline. With the
speaker recording placed first in every stacked file, so that role order is held
constant by construction, leave-one-dyad-out 5-NN accuracy is $0.766$ against a
majority-class level of $0.545$.
Taken alone this looks like a trait effect.

It is not stable under the choice of null. We ran two controls on the same
embedding and the same labels
(\tableref{tab:twonulls}). The \emph{frozen-embedding} control holds the trained
representation fixed and permutes AQ-magnitude labels across dyads 1000 times;
the \emph{retraining} control permutes complete dyad-level label signatures,
preserving class counts, and retrains CEBRA from scratch under each permutation
with identical preprocessing, configuration, seed, and training duration.

\begin{table}[htbp]
\floatconts
  {tab:twonulls}
  {\caption{Two permutation controls for dyad-level AQ magnitude, applied to the
  same data and the same labels. Empirical $p$ for the retraining control uses
  $(1+\#\{\mathrm{null}\ge\mathrm{real}\})/(1+B)$.}}
  {\begin{tabular}{lcccc}
  \hline
  Control & $B$ & Real 5-NN & Null mean $\pm$ sd & Empirical $p$\\
  \hline
  Frozen embedding, labels permuted & 1000 & 0.7588 & $0.4821 \pm 0.0340$ & 0.001\\
  Encoder retrained per permutation & 5    & 0.7657 & $0.7575 \pm 0.0104$ & 0.500\\
  \hline
  \end{tabular}}
\end{table}

The frozen control asks whether the geometry that was learned happens to contain
the label partition; the answer is affirmative whenever the encoder has learned
group identity, because the label is a coarse-graining of identity. The
retraining control asks whether comparable geometry arises from arbitrary
regroupings, and here it does: no real-run metric (accuracy, goodness of fit,
silhouette, spherical KS statistics, or mixture agreement) exceeded the
A six-class analysis and
one-versus-one SVM decoders reproduce the pattern
; pairwise separability there is
\emph{not} ordered by AQ distance, as a graded dyadic representation would
predict.

This does not establish that AQ magnitude has no neural correlate. It
establishes that a between-dyad labelling design cannot identify one
independently of dyad identity, and that the frozen-embedding null cannot
detect the problem.
\section{Manifold organization at individual level: role invariance and trait-level structure}
We next trained CEBRA on individual participant AQ, keeping speaker and listener recordings separate
(40 participants, 80 recordings, 6.08M samples; AQ 0--9; final loss 5.9917,
goodness of fit 0.3343). Unlike the dyad-level result, this one separates from an identity-aware null:
across 100 participant-label permutations with a matched 20{,}000-sample
subsample and dyad-level holdout, real decoding was $0.2470$ against a null mean
of $0.1157 \pm 0.0197$ ($p=0.0099$). Accuracy did not fall when the entire
co-recorded partner was also held out, exact-AQ accuracy rose from $0.2300$
under leave-one-participant-out to $0.2437$ under leave-one-dyad-out
(\tableref{tab:individual}), so the effect is not carried by within-dyad
leakage. It is nonetheless modest: exact-AQ exceeds the $0.2007$ majority
baseline by $\sim\!0.046$, balanced accuracy across ten levels is $0.188$, and
the AQ silhouette is $-0.167$. AQ grades the manifold without partitioning it.

\begin{table}[htbp]
\floatconts
  {tab:individual}
  {\caption{Participant-level AQ decoding with complete participants or complete
  dyads held out. Exact AQ uses ten levels; binary AQ splits at level~4.}}
  {\begin{tabular}{lcccc}
  \hline
  Holdout & Exact AQ & Exact bal. & Binary AQ & Binary bal.\\
  \hline
  Leave-one-participant-out & 0.2300 & 0.1705 & 0.6277 & 0.6165\\
  Leave-one-dyad-out & \textbf{0.2437} & \textbf{0.1882} & \textbf{0.6474} & \textbf{0.6374}\\
  \hline
  Majority baseline & 0.2007 & -- & 0.5740 & --\\
  Uniform chance & 0.1000 & -- & 0.5000 & --\\
  \hline
  \end{tabular}}
\end{table}

In the same embedding, speaker-versus-listener 5-NN accuracy under
leave-one-participant-out evaluation was $0.4736$ against a $0.5006$ majority
level, with silhouette $\approx 0$. Spherical Kolmogorov--Smirnov statistics for
role were small (0.013, 0.019, 0.021 for latitude, longitude, radius) relative
to the corresponding binary-AQ statistics (0.171, 0.224, 0.240)
(\figureref{fig:geometry}). Yet the two roles of one person are close on the
sphere: within-participant speaker--listener cosine similarity was
$0.9568 \pm 0.0564$ against $0.1776 \pm 0.6488$ across participants, and each
participant's alternate-role recording fell among the five most similar
recordings for 77.5\% of participants (12.5\% expected by chance). Role is thus
not represented as a state, but individual structure is preserved across it.

\begin{figure}[htbp]
\floatconts
  {fig:geometry}
  {\caption{Participant-level CEBRA geometry in longitude--latitude coordinates
  on $S^{2}$. AQ information is decodable across held-out participants, whereas
  speaker/listener role is at chance in the same representation.}}
  {%
    \subfigure[Coloured by participant AQ level.][c]{\label{fig:aq-lonlat}%
      \IfFileExists{figures/individual_aq_lonlat.png}{%
        \includegraphics[width=0.44\linewidth]{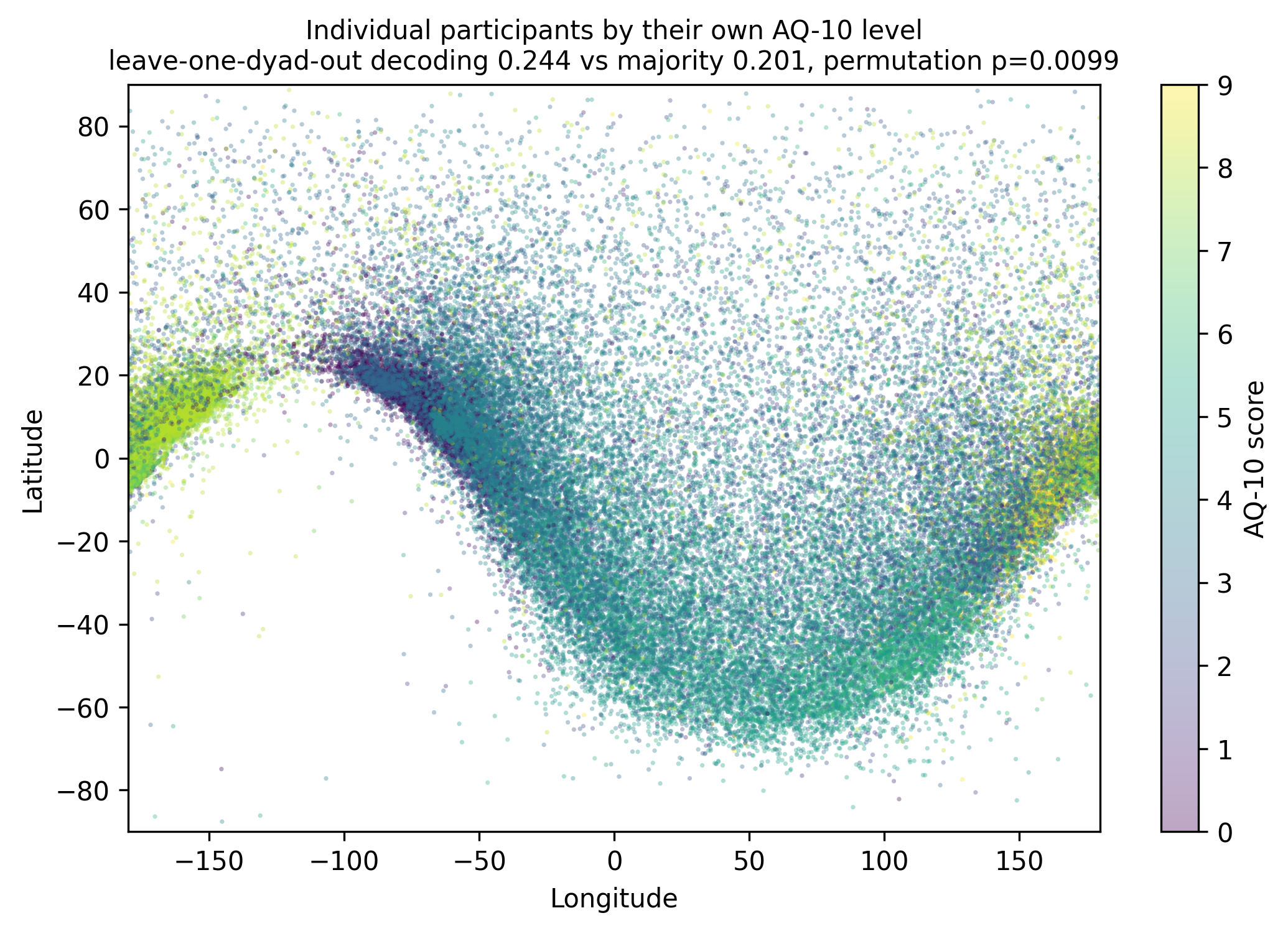}}{%
        \fbox{\parbox[c][1.25in][c]{0.40\linewidth}{\centering
          \texttt{individual\_aq\_lonlat.png}}}}}%
    \qquad
    \subfigure[Coloured by speaker/listener role.][c]{\label{fig:role-lonlat}%
      \IfFileExists{figures/individual_role_lonlat.png}{%
        \includegraphics[width=0.44\linewidth]{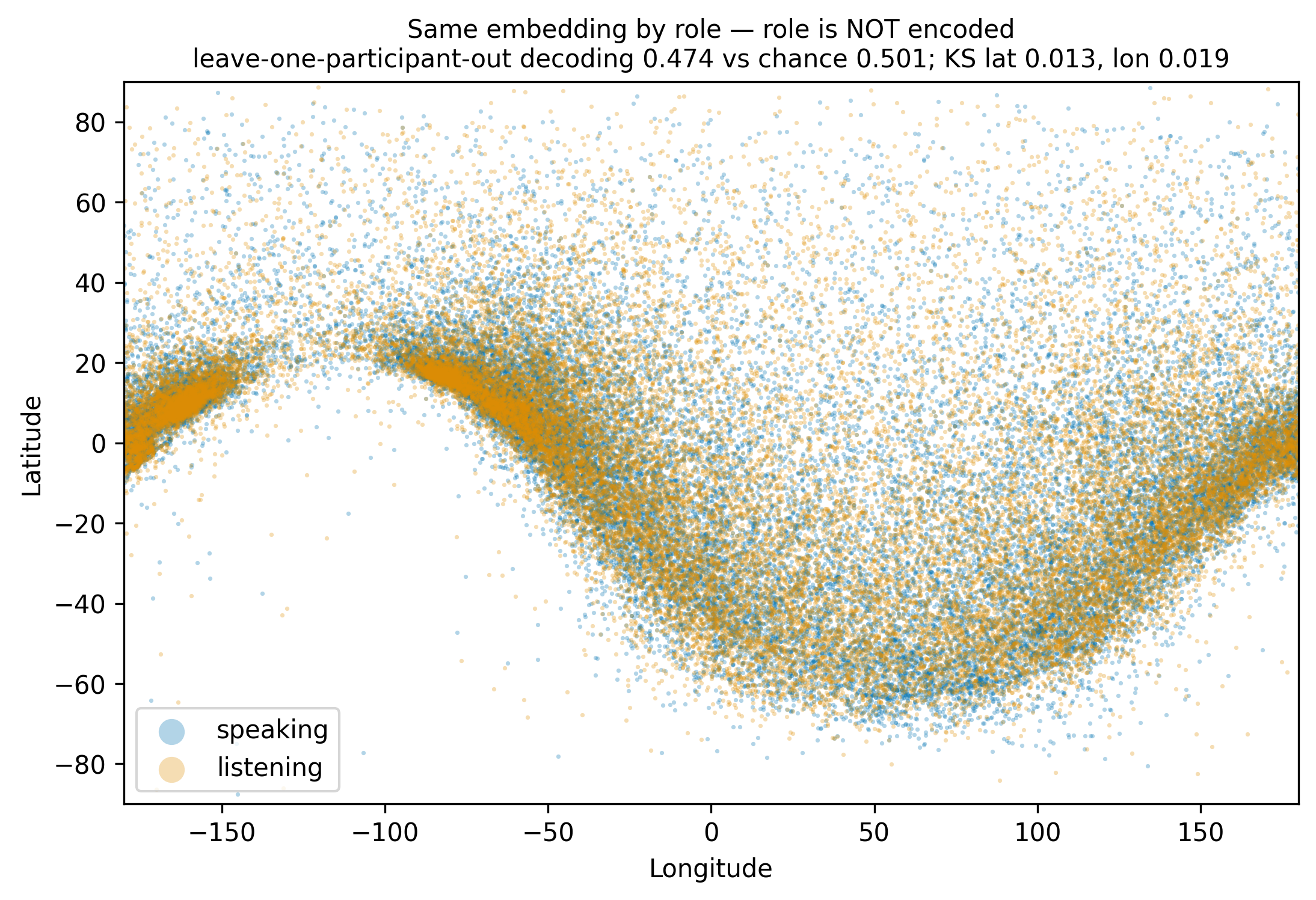}}{%
        \fbox{\parbox[c][1.25in][c]{0.40\linewidth}{\centering
          \texttt{individual\_role\_lonlat.png}}}}}%
  }
\end{figure}

\section{Discussion}
\label{sec:discussion}

The generalizeable result of embedding analysis is that decoding accuracy, computed on a contrastive embedding,
is not a test of the label when the label is constant within a recording
group. The frozen-embedding permutation essentially asks if the geometry that
was learned in contrastive embedding happens to contain the label partition; the answer will be affirmative
whenever the encoder has learned group identity, because the label is, in fact, a
coarse-grained equivalent of identity. The retraining permutation asks if
comparable representaitonal geometry arises from arbitrary regroupings. In our data, the two procedures differ by
orders of magnitude in $p$-value while operating on identical
embeddings and identical labels. Since grouped neural datasets are the norm
rather than the exception, we would expect this issue to occur widely wherever contrastive embeddings
are evaluated by decoding. 

Interestingly, participant-level AQ embedding
generalizes to fully held-out dyads and separates from an identity-aware
null, while speaker-versus-listener role is at chance in the same
embedding, indicating within-person recordings are highly similar across conversational roles.
The neural manifold, then, is primarilty organized by who is being recorded (individual neural dynamics and state space - cf. \citep{ahn2025distinct,malaia2020dysregulation,malaia2016altered, mccleod2024neurocomputational})
rather than by what they are doing. For dyadic neuroscience, this means the need to adopt experimental designs
in which the label of interest varies within a
recording group (either using within-dyad manipulations, or constructing repeated pairings across partners).

\paragraph{Limitations.}
Our sample (33 dyads at the dyad level; 40 participants at the participant level) was
over-represented for lower-AQ (neurotypical) participants, such that the classes with larger
$|\Delta\mathrm{AQ}|$ contained as few as three dyads. Thus, the
six-class analysis is sample-limited independently of any confound.

\paragraph{Broader impacts.}
Our approach advances understanding of interpersonal neural mechanisms in
autism and social communication more broadly. By moving beyond deficit-based
models to examine dyadic interactions, this research supports more nuanced
views of neurodiversity and could inform more effective,
relationship-focused interventions. Geometric representation approaches could contribute
to development of objective biomarkers for social communication challenges;
however, the present results are a demonstration of why such measures should not be treated as
biomarkers judged by the strength of decoding accuracy alone.

%
\acks{HH was a contributor in the 2026 Google Summer of Code Program with
Machine Learning for Science (ML4SCI) Organization. Data collection for this
research was supported by funding to EM from the University of Alabama
College of Arts and Sciences.}

\bibliography{refs}

 \end{document}